\documentclass[a4paper,11pt]{article}
\usepackage{pos}
\usepackage{CJKutf8}

\begin{document}

\title{Hadron Structure: Perspective and Insights}


\author[a]{Daniele Binosi%
    $\,^{\href{https://orcid.org/0000-0003-1742-4689}{\textcolor[rgb]{0.00,1.00,0.00}{\sf ID}}}$}

\author*[b,c]{Craig D.\ Roberts%
       $^{\href{https://orcid.org/0000-0002-2937-1361}{\textcolor[rgb]{0.00,1.00,0.00}{\sf ID}}}$}

\author[d,e]{Zhao-Qian Yao
(\hspace*{0.1em}\makebox[3.3em][l]{\hspace*{0em}\raisebox{-0.7ex}{\includegraphics[width=2.9em]{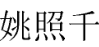}}}
\hspace*{-0.6em})
$\,^{\href{https://orcid.org/0000-0002-9621-6994}{\textcolor[rgb]{0.00,1.00,0.00}{\sf ID}}\,}$}

\affiliation[a]{European Centre for Theoretical Studies in Nuclear Physics
            and Related Areas  (\href{https://ror.org/01gzye136}{ECT*}), Villa Tambosi, Strada delle Tabarelle 286, I-38123 Villain (TN), Italy}

\affiliation[b]{School of Physics, \href{https://ror.org/01rxvg760}{Nanjing University},
Nanjing, Jiangsu 210093, China}

\affiliation[c]{Institute for Nonperturbative Physics, \href{https://ror.org/01rxvg760}{Nanjing University}, Nanjing, Jiangsu 210093, China}

\affiliation[d]{Department of Integrated Sciences and Center for Advanced Studies in Physics, Mathematics and Computation, \href{https://ror.org/03a1kt624}{University of Huelva}, E-21071 Huelva, Spain}

\affiliation[e]{Dpto. Sistemas F\'isicos, Qu\'imicos y Naturales,
\href{https://ror.org/02z749649}{Univ.\ Pablo de Olavide}, E-41013 Sevilla, Spain}

\emailAdd{cdroberts@nju.edu.cn}




\abstract{
\vspace*{-10ex}
\rightline{\sf NJU-INP 097/25}

\vspace*{9ex}
The bulk of visible mass is supposed to emerge from nonperturbative dynamics within quantum chromodynamics (QCD) -- the strong interaction sector of the Standard Model.  Following years of development and refinement, continuum and lattice Schwinger function methods have recently joined in revealing the three pillars that support this emergent hadron mass (EHM); namely, a nonzero gluon mass-scale, a process-independent effective charge, and dressed-quarks with constituent-like masses.  One may argue that EHM and confinement are inextricably linked; and theory is now working to expose their manifold expressions in hadron observables and highlight the types of measurements that can be made in order to validate the paradigm.  This contribution sketches the role played by EHM in shaping hadron electromagnetic and gravitational form factors, exciting nucleon resonances, and moulding hadron parton distributions.
}

\FullConference{The XVIth Quark Confinement and the Hadron Spectrum Conference (QCHSC24)\\
 19-24 August, 2024\\
 Cairns Convention Centre, Cairns, Queensland, Australia\\}


\maketitle

\section{Foundation}
It is worth beginning with a few basic questions in Nature that insightful physics might answer in the foreseeable future: (\emph{a}) What is the origin of the nuclear-physics mass-scale, $m_p$, the proton mass, which sets the characteristic size of all visible matter; (\emph{b}) Whatever it is, why is the pion, with its unnaturally small, lepton-like mass, $m_\pi \approx m_p/7 \approx m_{\rm muon}$, seemingly oblivious; and (\emph{c}) How is(are) the underlying mechanism(s) expressed in measurable quantities?  The answers to these questions should be objective, \emph{viz}.\ independent of reference frame, probe resolution scale, and other observer-dependent (subjective) considerations.  Moreover, concerning (\emph{c}), one should expect the expressions to be system specific; and that is a good thing, because it means that any theoretical framework, which pretends to deliver answers, can be exhaustively tested.  Finally, the answers are important because this nuclear physics mass scale emerged around 1$\mu$s after the Big Bang and subsequently had a critical influence on the evolution of the observed Universe.

There is one known mass generating mechanism in the Standard Model (SM); namely, Higgs boson couplings in the SM Lagrangian density.  This part is phenomenologically successful and understood at that level.  This means that although a Higgs boson like object has been found \cite{Aad:2012tfa, Chatrchyan:2012xdj}, there are deep theoretical problems with Higgs physics, \emph{e.g}., the hierarchy problem is real; quantum corrections to the mass of a scalar boson are quadratically divergent, \emph{viz}.\ not renormalisable; large quantum contributions to $m_{\rm Higgs}^2$ would inevitably make the mass huge; and avoiding such an outcome would require (incredible) fine-tuning cancellation between the quadratically divergent corrections and the Higgs bare mass.  A means of circumventing these and related issues is being sought, \emph{e.g}., realising the Higgs boson as a composite particle \cite{Cacciapaglia:2020kgq}.

\begin{figure}[t]
\centerline{%
\includegraphics[clip, width=0.66\textwidth]{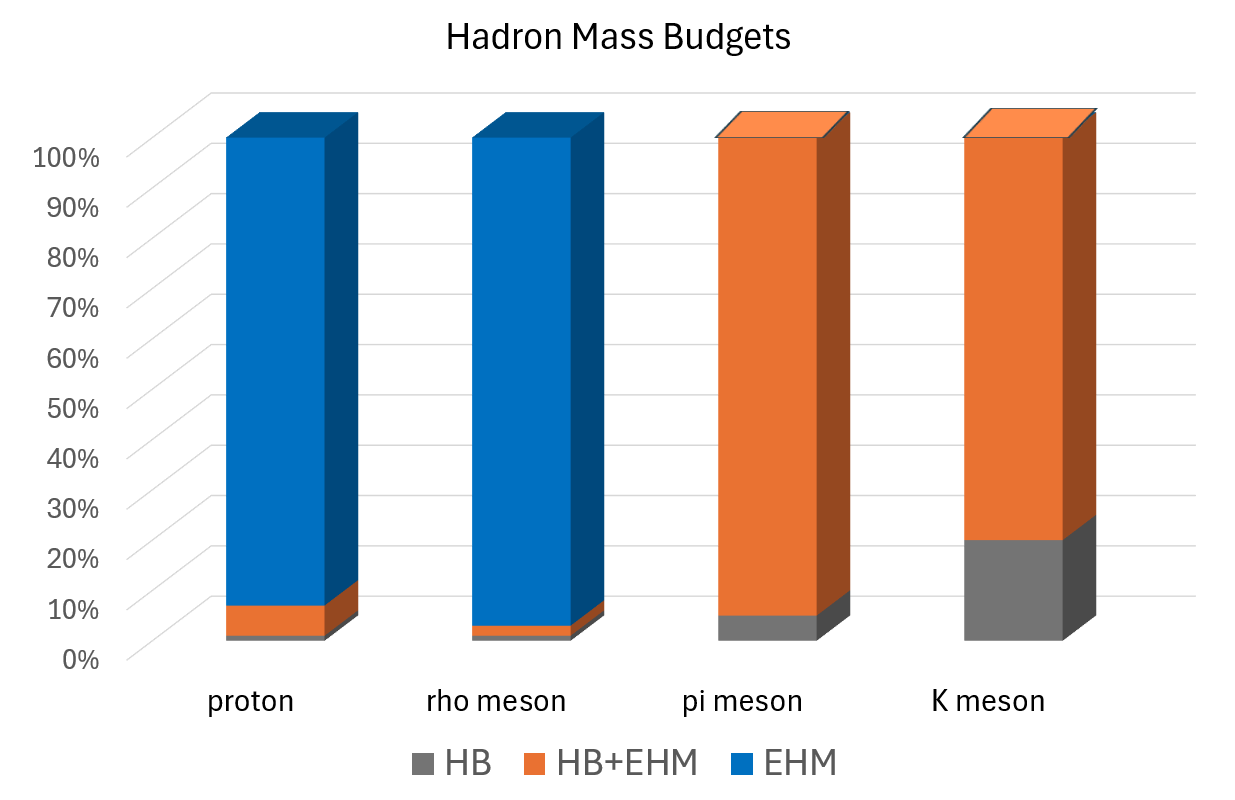}}
\caption{\label{MassBudgets}
Hadron Mass budgets.  The decompositions are gauge invariant, Poincar\'e invariant, and renormalisation scale independent.
Note the dramatic switch in the dominant contributions: two left bars ($p$, $\rho$) \emph{cf}.\ two right ($\pi$, $K$).  Particularly dramatic is the difference between $\rho$ and $\pi$ mass budgets.  From a quark model perspective, the $\rho$ is simply the pion's spin-flip partner: yet, their mass budgets are completely different.  (Additional details are available in Ref.\,\cite[Sec.\,2]{Ding:2022ows}.)
}
\end{figure}

Regarding QCD, Higgs boson couplings generate the quark current masses; and in the lighter quark sector, they lead to renormalisation point invariant scales of $\hat m_{u,d} \approx 0.006\,$GeV, $\hat m_s \approx 0.16\,$GeV.  Evidently, since the proton's valence quark structure is $u+u+d$, Higgs boson effects generate $< 2$\% of the proton mass -- see Fig.\,\ref{MassBudgets}.  So, Nature must have another, much more effective mechanism for generating everyday mass.  Exposing this mechanism is the search for the origin of emergent hadron mass (EHM) \cite{Roberts:2021nhw, Binosi:2022djx, Ding:2022ows, Roberts:2022rxm, Ferreira:2023fva, Carman:2023zke, Raya:2024ejx}.

Figure~\ref{MassBudgets} depicts a breakdown of some hadron masses into three distinct components: HB -- that owing to the Higgs boson alone; HB + EHM -- the constructive interference contribution between Higgs boson and EHM effects; and EHM -- that part which is a pure expression of emergent hadron mass.
Evidently, for the proton and $\rho$-meson, EHM is the overwhelmingly dominant component: it is responsible for $\approx 95$\% of the measured mass.
Notably, from a quark model perspective, for instance, the $\rho$-meson is merely the pion's spin-flip partner.  Yet, their mass budgets are very different: the pion mass receives zero EHM-only contribution and 95\% is generated by HB+EHM interference.
This is because, in QCD, the pion is a (pseudo) Nambu-Goldstone boson, \emph{i.e}., much more than simply a spin-flip partner of the $\rho$.  Consequently, owing to fundamental quark-level Goldberger-Treiman relations \cite{Maris:1997hd, Qin:2014vya}, the large positive EHM-only contributions to the two one-body (valence quark) dressings are precisely cancelled by negative EHM corrections to the two-body binding energy \cite{Roberts:2016vyn}.  This leaves HB+EHM as the leading, nonzero component, which is the outcome expressed by the Gell-Mann--Oakes--Renner relation \cite{GellMann:1968rz, Chang:2011mu}.

The kaon mass budget is also drawn in Fig.\,\ref{MassBudgets}.  Absent Higgs boson couplings, \emph{i.e}., in the chiral limit, the $K$-meson is a Nambu-Goldstone boson.  In fact, it is indistinguishable from the pion.  On the other hand, with realistic Higgs boson couplings, $\hat m_s \approx 27 (\hat m_u+ \hat m_d)/2$ \cite{ParticleDataGroup:2024cfk}; hence, the HB portion of the $K$ mass budget accounts for 20\% of $m_K$.  The remainder (80\%) is produced by HB+EHM interference.  There is still no EHM-only component.  Consequently, comparisons between $\pi$ and $K$ properties provide excellent opportunities for studying Higgs boson modulation of EHM, because the HB mass fraction is four-times larger in kaons than in pions.  Furthermore, Fig.\,\ref{MassBudgets} highlights that complementary  information can be obtained from comparisons between baryons/vector mesons and the set of kindred pseudoscalar mesons.  Some of these opportunities are reviewed elsewhere \cite{Roberts:2021nhw, Ding:2022ows}.

Having seen its various impacts on hadron masses, one is pressed to ask, \emph{inter alia}:
What is the origin of EHM;
Does it lie within QCD;
What are EHM's connections (if any) with gluon and quark confinement, dynamical chiral symmetry breaking, and the appearance and properties of pions and kaons -- Nature's most fundamental Nambu-Goldstone bosons;
and How does the Higgs boson mass generating mechanism modulate and influence the expressions of EHM in observables?
The sum of these questions amounts to the following: What is EHM, where does it come from, and what are its observable effects?

Supposing EHM is a SM feature, then it should lie within the QCD Lagrangian density:
\begin{subequations}
\label{QCDdefine}
\begin{align}
{\mathscr L}_{\rm QCD} & = \sum_{{\mathscr f}=u,d,s,\ldots}
\bar{q}_{\mathscr f} [\gamma\cdot\partial
 + i g \tfrac{1}{2} \lambda^a\gamma\cdot A^a+ m_{\mathscr f}] q_{\mathscr f}
 + \tfrac{1}{4} G^a_{\mu\nu} G^a_{\mu\nu},\\
%
%
\label{gluonSI}
G^a_{\mu\nu} & = \partial_\mu A^a_\nu + \partial_\nu A^a_\mu -
g f^{abc}A^b_\mu A^c_\nu,
\end{align}
\end{subequations}
where the fields $\{q_{\mathscr f}\,|\,{\mathscr f}=u,d,s,c,b,t\}$ are associated with the six known flavours of quarks; their current-masses, $\{m_{\mathscr f}\}$, are generated by the Higgs boson;
the gluon fields are $\{A_\mu^a\,|\,a=1,\ldots,8\}$, whose matrix structure is encoded in $\{\tfrac{1}{2}\lambda^a\}$, the generators of SU$(3)$ in the fundamental representation; and $g$ is the unique QCD coupling, using which one defines $\alpha_s = g^2/[4\pi]$.
There is no Lagrangian mass term for the gluons as that would violate gauge invariance.
(For simplicity, ghost fields are suppressed. See, \emph{e.g}., Ref.\,\cite{Pascual:1984zb} for more details.)

The density in Eq.\,\eqref{QCDdefine} was first written in Ref.\,\cite{Fritzsch:1972jv}.
It was discussed further in Ref.\,\cite{Fritzsch:1973pi}, which contains the following introductory remarks:
``\emph{The quarks come in three `colors,' but all physical states and interactions are supposed to be singlets with respect to the SU$(3)$ of color. Thus, we do not accept theories in which quarks are real, observable particles; nor do we allow any scheme in which the color non-singlet degrees of freedom can be excited.  Color is a perfect symmetry.}''   These remarks signal the key issue with QCD; namely, the degrees of freedom used to express the QCD Lagrangian are not those measured in detectors.  In attempting to solve QCD, theory is therefore confronted with the following questions:
(\emph{i}) What are the (asymptotic) detectable degrees-of-freedom;
(\emph{ii}) How are they built from the Lagrangian degrees-of-freedom;
(\emph{iii}) Is QCD really the theory of strong interactions;
and (\emph{iv}) Is QCD really a theory or just another effective field theory (EFT)?
Questions (\emph{iii}) and (\emph{iv}) overlap: even if QCD is only an EFT, valid on some large but nevertheless limited energy domain, it may nevertheless describe all that one might expect of a SM theory of strong interactions.  On the other hand, if QCD is really a theory, \emph{i.e}., a mathematically well-defined four-dimensional quantum gauge field theory, then we would be working with something unique, whose solution could have implications far beyond the SM.

\section{Genesis}
\label{SecGenesis}
The only thing that really makes a difference between QCD and quantum electrodynamics (QED) is the ``$g f^{abc}A^b_\mu A^c_\nu$'' term in Eq.\,\eqref{gluonSI}; and the difference it makes is huge.  This term means that gluons interact with each other at leading perturbative order in the analysis of any strong interaction observable.  For photons, on the other hand, self-interactions are a quantum loop effect, suppressed by four powers of the fine structure constant \cite{ATLAS:2019azn}.  Therefore, following the first appearance of ${\mathscr L}_{\rm QCD}$, it was soon demonstrated that the momentum evolution of the QCD coupling must exhibit the opposite pattern to that in QED, \emph{i.e}., that QCD is asymptotically free \cite{Politzer:2005kc, Wilczek:2005az, Gross:2005kv}: the closer coloured objects are to each other, the weaker is the strong interaction between them.

The flip-side of asymptotic freedom is that the coupling must grow with interparticle separation.  This is the seed for confinement because it means that long-wavelength gluons are strongly interacting; so, there are potentially enormous nonperturbative feedback effects in any solution of the gluon gap equation.  Appreciating this, Ref.\,\cite{Cornwall:1981zr} argued that ${\mathscr L}_{\rm QCD}$ must support the formation of gluon quasiparticles, each built from a countable infinity of massless gluon partons, and each being described by a momentum-dependent mass function that is large at infrared momenta: $\mu_G :=m_G(k^2 \simeq 0) = (0.5 \pm 0.2)\,$GeV.  This scale is half the proton mass!

With the dynamical generation of a gluon mass, one sees mass emerging from nothing.  An interacting theory, written in terms of massless gluon fields, produces dressed gluon fields which are characterised by a running mass that is large at infrared momenta.  This is the clearest expression of the QCD trace anomaly.  No symmetries are broken by this dynamical effect \cite{Schwinger:1962tn}.  Indeed, the QCD outcome is driven by qualitatively the same mechanism that was first exposed in that early work -- see Refs.\,\cite{Binosi:2022djx, Ferreira:2023fva, Deur:2023dzc} for contemporary perspectives.  As stressed therein, the emergence of $\mu_G$ is a QCD fact, revealed by both continuum and lattice studies of the two-point gluon Schwinger function (propagator): impossible to see in perturbation theory, it may nevertheless be the key to understanding the stability of QCD.

The emergence $\mu_G$ is communicated dynamically into the QCD running coupling, $\alpha_s(k^2)$.  Perturbatively, the definition of $\alpha_s(k^2)$ is unambiguous and its value is known with a precision of $\lesssim 1$\% \cite{Deur:2023dzc}.  To put this in context, the perturbative QED running coupling is known with precision better than $1/10^9$.  Nonperturbatively, on the other hand, the QED coupling is undefined -- see, \emph{e.g}., Refs.\,\cite{Rakow:1990jv, Gockeler:1994ci, Reenders:1999bg, Kizilersu:2014ela}; and different practitioners employ distinct definitions of $\alpha_s(k^2)$ \cite{Deur:2023dzc}.  At heart, the ambiguity in QCD arises because there are 8 relevant renormalisation constants in the theory; consequently, \emph{prima facie}, no one Schwinger function or momentum flow into that function is a better choice than any other.

\begin{figure}[t]
\centerline{\includegraphics[clip, width=0.66\textwidth]{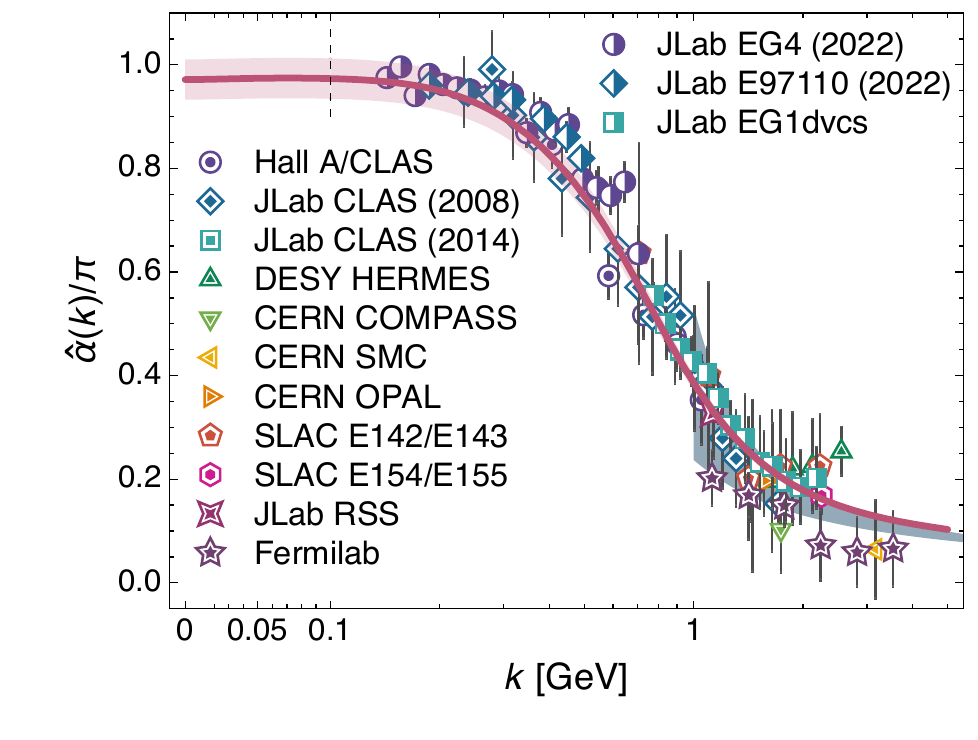}}
\caption{\label{Falpha}
Process-independent effective charge, $\hat{\alpha}(k)$, obtained by combining the best available results from continuum and lattice analyses of QCD's gauge sector \cite{Cui:2019dwv}.
%
%
Existing data on the process-dependent charge $\alpha_{g_1}$ \cite{Deur:2022msf, Deur:2023dzc}, defined via the Bjorken sum rule, are also shown -- see Ref.\,\cite[Fig.\,3]{Ding:2022ows} for the sources.
%
}
\end{figure}

The challenge of ambiguity was overcome following a realisation \cite{Binosi:2016nme} that, by using the pinch technique \cite{Pilaftsis:1996fh, Binosi:2009qm, Cornwall:2010upa} and background field method \cite{Abbott:1981ke}, one can define and calculate a unique, process-independent (PI) and renormalisation group invariant QCD analogue of the Gell-Mann--Low effective charge, which is the QED-given archetype of such charges.  Denoted $\hat\alpha(k^2)$, this QCD effective charge is obtained from a modified gluon vacuum polarisation.  As such, like in QED, only one momentum variable is involved and the coupling is precisely the same, independent of the scattering process one considers: in QCD, this means
gluon+gluon $\to$ gluon+gluon, quark+quark $\to$ quark+quark, etc.
Today, the most precise determination of $\hat\alpha(k^2)$ is described in Ref.\,\cite{Cui:2019dwv}, which obtained a parameter-free prediction by combining results from continuum analyses of QCD's gauge sector and lQCD configurations generated with three domain-wall fermions at the physical pion mass \cite{Blum:2014tka, Boyle:2015exm, Boyle:2017jwu}. The resulting charge is drawn in Fig.\,\ref{Falpha}.

The PI charge has a range of important properties.  Here, it is worth highlighting two.
First, whereas any perturbatively defined coupling in QCD will exhibit a Landau pole, this is absent in $\hat\alpha(k^2)$: it is eliminated by emergence of the gluon mass scale.
Gluons provide antiscreening in QCD, \emph{i.e}., gluon loops cause the coupling to grow as $k^2$ becomes smaller.  However, loops with massive gluons are suppressed when the momentum exchanged lies below $\mu_G$.  So, antiscreening is blocked.
Similarly, since EHM also provides light-quarks with a running mass that is large at infrared momenta -- see, \emph{e.g}., Ref.\,\cite[Fig.\,2.5]{Roberts:2021nhw}, then quark screening is blocked on the same domain.
These outcomes eliminate those dynamical effects which are responsible for the coupling's running; consequently, the coupling stops running and QCD becomes a practically conformal theory at infrared momenta.
The PI charge saturates at a value $\hat\alpha(k^2\simeq 0) = 0.97(4)$.

Second, the process-dependent charge defined via the Bjorken sum rule \cite{Deur:2022msf, Deur:2023dzc}, $\alpha_{g_1}(k^2)$, is almost indistinguishable from the predicted PI charge.  On any domain within which perturbation theory is valid, the reason for this near equality is obvious.
Namely, on $k^2 \gg (m_p/2)^2$,
$\alpha_{g_1}(k^2)/\hat\alpha(k^2) \approx 1+\frac{1}{20} \alpha_{\overline{\rm MS}}(k^2)$, where $\alpha_{\overline{\rm MS}}$ is the standard running coupling.  At the charm quark current mass, the ratio is $1.007$, \emph{i.e}., practically unity.
In the far infrared, on the other hand, the Bjorken charge saturates to $\alpha_{g_1}(k^2=0)=\pi$; hence, on $k^2 \ll (m_p/2)^2$, $\alpha_{g_1}(k^2)/\hat\alpha(k^2) = 1.03(4)$.
At infrared momenta, this near identity may be explained by the fact that the Bjorken sum rule is an isospin non-singlet relation.  Consequently, many dynamical contributions that might distinguish between the two charges are eliminated.
These remarks explain the result displayed in Fig.\,\ref{Falpha}, \emph{viz}.\ for practical intents and purposes, the process-dependent Bjorken charge is indistinguishable from QCD's PI charge \cite{Binosi:2016nme, Cui:2019dwv}.  An accessible, broader perspective is supplied elsewhere \cite{Brodsky:2024zev}.

It is now worth providing a short summary.
Absent Higgs boson couplings, the QCD Lagrangian is scale invariant; hence, naively, it should not support massive particles of any kind \cite{Roberts:2016vyn}.
Notwithstanding that, massless gluon partons transmogrify into gluon quasiparticles, which are characterised by a proton-size infrared mass scale.
As a consequence, QCD produces a momentum dependent effective charge that saturates at infrared momenta, is finite for all spacelike momenta, $k^2>0$, and falls monotonically away from its global maximum with increasing $k^2$.
Embedding these elements in the quark gap equation, one obtains a solution that involves a running quark mass, $M(k^2)$, which is also large at infrared momenta, possessing a scale that readily explains $m_p$.

The issue of gauge invariance is sometimes raised in connection with the material presented above, but any related concerns are baseless.
Continuum calculations are typically performed in Landau gauge for three practical reasons.
(\emph{i}) Landau gauge is a fixed point of the QCD renormalisation group; so, the gauge parameter does not run and, once fixed, Landau gauge remains Landau gauge.
(\emph{ii}) In delivering manifestly Poincar\'e covariant $n$-point functions, Landau gauge ensures Poincar\'e invariant results for all calculated observables.
(\emph{iii}) Landau gauge can readily be fixed in studies employing lattice-regularised QCD (lQCD); so, valid comparisons can straightforwardly be made between results obtained using these two complementary nonperturbative approaches.  Few other gauges are accessible using lQCD.
Point (\emph{iii}) is significant because lQCD is a manifestly gauge invariant approach.  Thus, agreement between the Landau gauge continuum prediction for a given $n$-point function and the Landau gauge projection of lQCD simulations that yield the same Schwinger function is practical confirmation of the gauge covariance of the continuum result.

It is worth adding that solutions of QCD's quantum equations of motions, \emph{e.g}., Dyson-Schwinger equations, are gauge covariant.
Gauge transforms are nugatory, in the sense that they introduce non-dynamical transformations of gauge covariant objects.
Consequently, such quantities, properly combined into a gauge invariant matrix element, will deliver gauge invariant results.
These remarks are confirmed, in practice, by comparisons between continuum predictions and experimental observables because the measured quantities are plainly gauge invariant and an approach that directly connects elementary $n$-point Schwinger functions with measured quantities must be preserving gauge invariance if it is delivering agreement with the bulk of those observables.
These last remarks entail that all features of the Landau-gauge elementary $n$-point functions are expressed in calculated observables.
Thus, for instance, whilst the nonperturbative running of the dressed quark mass may not itself be directly observable, its impacts on observables are -- see, \emph{e.g}., Ref.\,\cite{Carman:2023zke}.


The three pillars of EHM are expressed in every strong interaction observable.
Now, theory is challenged to elucidate their observable consequences and identify paths to measuring them; and experiment is charged with testing those predictions.
Validating the EHM paradigm has the potential to answer whether QCD is really a theory and, thereby, determine the boundaries of the SM.  This is critical because one cannot build a Theory of Everything before it is known whether QCD must be a part of it.
An array of operating, planned, and desired high-luminosity, high-energy facilities should make this possible
\cite{Denisov:2018unj, Aguilar:2019teb, Brodsky:2020vco, Chen:2020ijn, Anderle:2021wcy, Arrington:2021biu, Aoki:2021cqa, Quintans:2022utc, Wang:2022xad}.

\section{Aspects of Nucleon Structure}
The potential efficacy of \emph{ab initio} continuum calculations of hadron observables via matrix elements built with elementary Schwinger functions was demonstrated in Refs.\,\cite{Maris:1997tm, Maris:2000sk, Eichmann:2009qa, Eichmann:2011vu}.  Today, following refinement of those early studies, a diverse, unified set of parameter-free, empirically benchmarked predictions are available.  They include results for pion, kaon, nucleon elastic electromagnetic and gravitational form factors and their species decompositions \cite{Yao:2024drm, Xu:2023izo, Yao:2024uej, Yao:2024ixu}, and $\pi$ and $K$ structure functions \cite{Cui:2020tdf}, all with direct connections to the three pillars of emergent hadron mass.

\begin{figure}[t]
\centerline{%
\includegraphics[clip, width=0.66\linewidth]{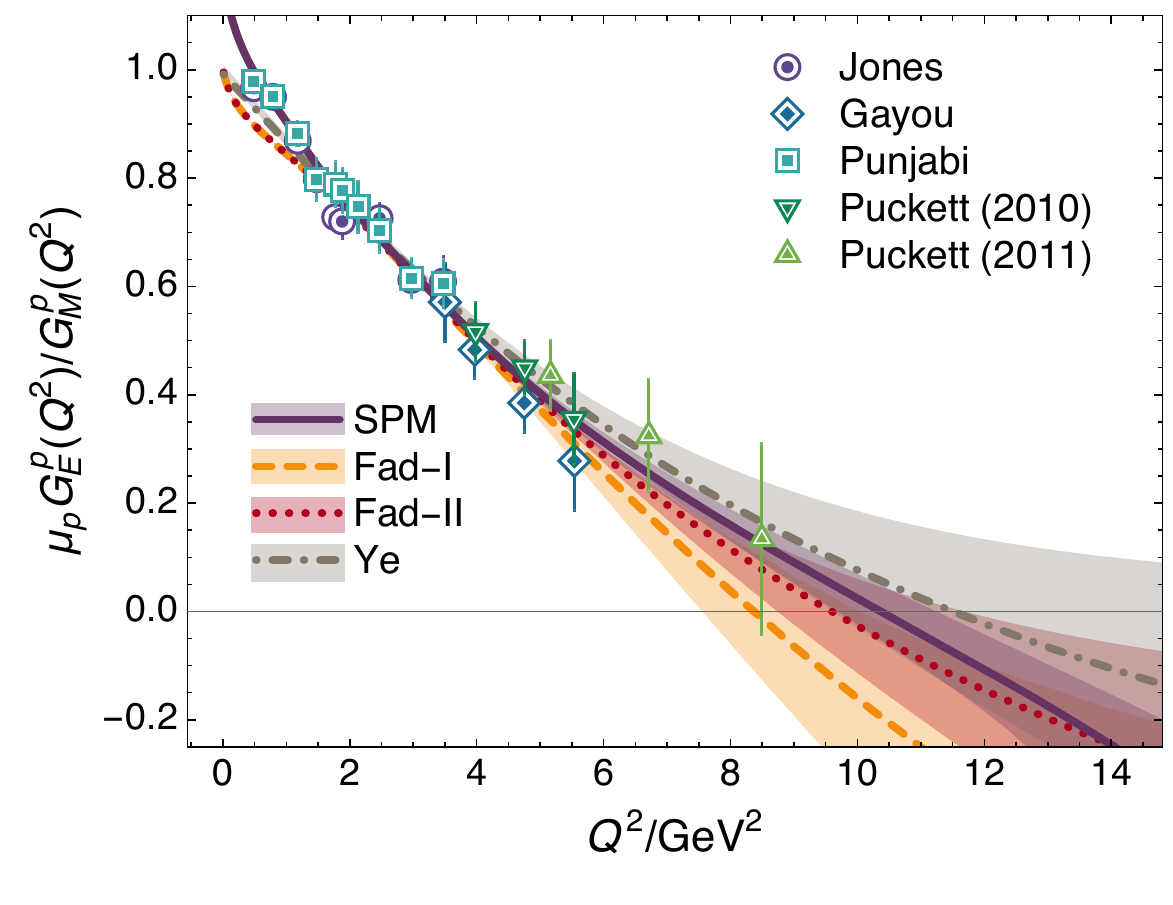}}
\caption{\label{SPMcfYe}
Prediction for the ratio $\mu_p G_E^p(Q^2)/ G_M^p(Q^2)$ obtained in the objective SPM analysis of available data \cite{Jones:1999rz, Gayou:2001qd, Punjabi:2005wq, Puckett:2010ac, Puckett:2017flj} that is described in Ref.\,\cite{Cheng:2024cxk}.
For comparison, the figure also depicts the parameter-free Faddeev equation predictions \cite[Fad-I, Fad-II]{Yao:2024uej} and the curve obtained via a subjective phenomenological fit to the world's electron + nucleon scattering data \cite[Ye]{Ye:2017gyb}.
Both theory and phenomenology deliver a zero crossing at a location that is compatible with the SPM prediction.
}
\end{figure}

Given these new results obtained using continuum Schwinger function methods (CSMs), it is worth placing them in context with a highlight of 21$^{\rm st}$ century experiment; namely, the collection of data  \cite{Jones:1999rz, Gayou:2001qd, Punjabi:2005wq, Puckett:2010ac, Puckett:2017flj} that hints at existence of a zero in $G_E^p$, the proton elastic electric form factor -- see Fig.\,\ref{SPMcfYe}.
The Faddeev equation studies in Ref.\,\cite{Yao:2024uej} predict that $G_E^p$ exhibits a zero at:
\begin{equation}
Q^2 = 8.86_{-0.86}^{+1.92} {\rm GeV}^2.
\end{equation}
On the other hand, the neutron electric form factor, $G_E^n$, is positive definite.  Consequently,
\begin{equation}
  G_E^n(Q^2) > G_E^p(Q^2) \; \mbox{on} \; Q^2 \geq 4.7\,\mbox{\rm GeV}^2 ,
\end{equation}
\emph{viz}.\ on this domain, somewhat against intuition, the electric form factor of the charge-neutral neutron is larger than that of the charge-one proton.

Verification of these predictions is within the reach of experiments either underway at JLab\,12 or currently being analysed.  Meanwhile, working with available data \cite{Jones:1999rz, Gayou:2001qd, Punjabi:2005wq, Puckett:2010ac, Puckett:2017flj} on the ratio of proton electric and magnetic form factors, $\mu_p G_E^p(Q^2)/ G_M^p(Q^2)$, the Schlessinger point method (SPM) \cite{Schlessinger:1966zz, PhysRev.167.1411, Tripolt:2016cya} was used in Ref.\,\cite{Cheng:2024cxk} to objectively assess the likelihood that the data support the existence of a zero in $G_E^p(Q^2)$ and, if so, its probable location.  The analysis reveals that, with 50\% confidence, the data are consistent with the existence of a zero in the ratio on $Q^2 \leq 10.37\,$GeV$^2$.  The confidence level increases to $99.9$\% on $Q^2 \leq 13.06\,$GeV$^2$.  Significantly, the likelihood that existing data are consistent with no zero in the ratio on $Q^2 \leq 14.49\,$GeV$^2$ is $1/1$-million.  This outcome challenges the viability of many models and calculations -- see, \emph{e.g}., Refs.\,\cite{Anikin:2013aka, Giannini:2015zia, Sufian:2016hwn, Kallidonis:2018cas, Xu:2021wwj}.
Notably, the data fit in Ref.\,\cite[Punjabi]{Punjabi:2015bba} locates a zero at $Q^2 \simeq 13.1\,$GeV$^2$ and the earlier fit in Ref.\,\cite[Kelly]{Kelly:2004hm} at $Q^2$ near 15\,GeV$^2$, which is marginally consistent with the SPM prediction.

It is worth recording some remarks on the physical significance of a zero in $G_E^p(Q^2)$.  Certainly, relativistic effects are important in describing available $\mu_p G_E^p(Q^2)/ G_M^p(Q^2)$ data; yet, particular features of QCD may be equally or more significant.
For instance, as discussed above, owing to dynamical chiral symmetry breaking, a corollary of emergent hadron mass (EHM), light quarks acquire a strongly momentum dependent running mass that is large at infrared momenta \cite{Binosi:2016wcx}: $M(k^2=0) \approx 0.35\,$GeV.
Quark + interacting-diquark Faddeev equation descriptions of proton structure \cite{Barabanov:2020jvn} -- discussed further in Sec.\,\ref{SecResonance} -- suggest that the rate at which $M(k^2)$ runs toward its ultraviolet ($k^2/m_p^2 \gg 1$) current-mass limit has a material influence on the proton Pauli form factor \cite{Wilson:2011aa, Cloet:2013gva}: if the evolution is very rapid, \emph{i.e}., perturbative physics is quickly recovered, then $\mu_p G_E^p(Q^2)/ G_M^p(Q^2)$ does not exhibit a zero, whereas a zero is found with a slower transition from the nonperturbative to the perturbative domain.  This is an expression of the remarks made in the penultimate paragraph of Sec.\,\ref{SecGenesis}, \emph{viz}.\ whilst $M(k^2)$ may not itself be directly observable, its impact on observables is.  The mass function obtained as part of the set of solutions necessary to produce the Faddeev equation results drawn in Fig.\,\ref{SPMcfYe} matches that found in QCD -- see Ref.\,\cite[Fig.\,S.5]{Yao:2024ixu}.

Employing the same continuum approach to nucleon properties, Ref.\,\cite{Yao:2024ixu} delivered parameter-free predictions for all three nonzero nucleon gravitational form factors that characterise the expectation value of the QCD energy-momentum tensor in the nucleon:
\begin{align}
m_N \Lambda_{\mu\nu}^{Ng}(Q) & = - \Lambda_+(p_f)
[
K_\mu K_\nu A(Q^2)  \nonumber \\
& \quad + i K_{\left\{\mu\right.}\!\sigma_{\left.\nu\right\}}\,\!_\rho Q_\rho J(Q^2)  + \tfrac{1}{4} (Q_\mu Q_\nu - \delta_{\mu\nu} Q^2) D(Q^2)
]
\Lambda_+(p_i)  \,,
\label{EMTproton}
\end{align}
where
$p_{i,f}$ are the momenta of the incoming/outgoing nucleon, $p_{i,f}^2=-m_N^2$,
$K=(p_i+p_f)/2$, $Q=p_f-p_i$;
all Dirac matrices are standard \cite[Sec.\,2]{Roberts:2000aa}, with $\sigma_{\mu\nu}= (i/2)[\gamma_\mu,\gamma_\nu]$;
$\Lambda_+$ is the projection operator that delivers a positive energy nucleon;
and $a_{\left\{\mu\right.}\!b_{\left.\nu\right\}}=(a_\mu b_\nu + a_\nu b_\mu)/2$.
Importantly, each of the form factors in Eq.\,\eqref{EMTproton} is Poincar\'e invariant; hence, observable.

In Eq.\,\eqref{EMTproton}, $A$ is the nucleon mass distribution form factor and the other form factors relate to the distribution of total angular momentum, $J$, and in-nucleon pressure and shear forces, $D$.
In the forward limit, $Q^2=0$, symmetries entail $A(0)=1$, $J(0)=1/2$.
$D(0)$ is also a conserved charge but, like the nucleon axial charge, $g_A$, its value is a dynamical property.
It has been described as the last unknown global property of the nucleon \cite{Polyakov:2018zvc}; thus, is a focus of much attention.  The CSM prediction is \cite{Yao:2024ixu}: $D(0) = -3.11(1)$.  This value is confirmed by the data-informed extraction in Ref.\,\cite{Cao:2024zlf}: $D(0) = -3.38_{-0.32}^{+0.26}$.
For the pion, the analogous value is \cite{Xu:2023izo}: $[-\theta_1(Q^2=0)] = -0.97$.

Using the form factors in Eq.\,\eqref{EMTproton}, one may construct the proton mass-energy density form factor
\begin{equation}
{\cal M}(Q^2) =
A(Q^2) + \frac{Q^2}{4 m_p^2} [A(Q^2)-2 J(Q^2)+D(Q^2)]\,.
\label{MQ2}
\end{equation}
Insofar as the CSM analysis is concerned, the prediction for ${\cal M}(Q^2)$ is positive definite on a domain that extends to (at least) $Q^2=100\,$GeV$^2$.

In the Breit-Frame, ${\cal M}(Q^2) $ delivers a spatial density which is a direct analogue of the nucleon electric charge distribution that is measured in electron scattering.
Notably, since the source function is Poincar\'e-invariant, then whether one builds a spatial density using a three- or two-dimensional Fourier transform is immaterial.  No projective mapping can add new information to the Poincar\'e-invariant source function and any interpretation of the mapping's outcome will always be practitioner dependent.
As usual, the associated mass-energy radius may be obtained via
\begin{equation}
\langle r^2\rangle_{\rm mass} = -6 \left. \frac{d}{dQ^2}\ln {\cal M}(Q^2) \right|_{Q^2=0}
= -6 \left. \frac{d}{dQ^2} A(Q^2) \right|_{Q^2=0}
- \frac{3}{2m_p^2}D(0)\,.
\end{equation}
The similarity with the expression for the proton electric charge radius is plain.  In both cases, the radius increases as the magnitude of the symmetry-unconstrained pressure/anomalous-magnetic-moment term increases and with the same rate.

A normal force distribution form factor may also be defined.  The associated ``mechanical'' radius is obtained via
\begin{equation}
\langle r^2\rangle_{\rm mech} = \frac{6}{\int_0^\infty dt\,[D(t=Q^2)/D(0)]}\,.
\end{equation}
The CSM analysis predicts \cite{Yao:2024ixu}:
$r_{\rm mass} = 0.81(5) r_{\rm ch} > r_{\rm mech} = 0.72(2) r_{\rm ch}$,
where $r_{\rm ch}=0.887(3)\,$fm is the proton charge radius calculated using the same framework \cite{Yao:2024uej}.

In contrast to the Poincar\'e-invariant measurable form factors discussed above, any species decomposition of these quantities is frame- and scale-dependent; hence, subjective.  At the hadron scale, $\zeta_{\cal H}$, the behaviour of each form factor is completely determined by that of the dressed valence quark degrees of freedom.  All glue and sea-quark contributions are sublimated into the dressed valence quarks.  This is simply a definition of dressing and is expressed in any QCD Schwinger function -- consider, \textit{e.g}., a diagrammatic expansion of the quark gap equation.

Pursuing a species decomposition via the all-orders (AO) evolution scheme \cite{Yin:2023dbw},  one finds that, for each form factor, the scale-dependent gluon:total-quark contribution ratio is a fixed number (constant, independent of $Q^2$).  At $\zeta=\zeta_2:=2\,$GeV, the value is:
\begin{equation}
{\mathscr g}(Q^2)/{\mathscr q}(Q^2) = 0.71(4)\,.
\label{gonqconstant}
\end{equation}
It is important to test this prediction because it entails that no species decomposition contains new information: everything that can be known is already contained in the scale-invariant observable form factors.  For instance, the relative contributions of gluons and quarks to the proton mass radius are just given by
$\langle r^2\rangle_{\mathscr p = \mathscr g, \mathscr q}^\zeta = \langle x \rangle_{\mathscr p}^\zeta r_{\rm mass}^2$, \emph{i.e}., the products of their scale-dependent light-front momentum fractions with the observable radius.
At present, the only comparison computations available are those provided by the lQCD results in Ref.\,\cite{Hackett:2023rif}, whose statistical uncertainties are large and whose systematic errors have neither been quantified nor controlled.  Within this uncertainty context, the lQCD results are consistent with the Eq.\,\eqref{gonqconstant} prediction: ${\mathscr g}/{\mathscr q}_{\mbox{lQCD}} = 0.82(18)$ -- see Ref.\,\cite[Fig.\,S.8]{Yao:2024ixu}.

In the nucleon, pressure and shear-force density profiles are determined by $D(Q^2)$.  This is why it is called the pressure (\emph{Druck}) form factor.
The CSM predictions for such profiles are drawn in Ref.\,\cite[Fig.\,3]{Yao:2024ixu}.  Reviewing those predictions, one finds that the near-core pressure in the pion is roughly twice that in the proton and the in-proton near-core pressure is an order of magnitude greater than that in a neutron star \cite{Ozel:2016oaf}: these hadrons, an uncountable number of which are within us all, are the densest known systems in the Universe, excepting black holes.

It is worth stressing that interpretations of the pressure distributions obtained from $D(Q^2)$ are analogous to those associated with the distributions described by the mass and spin form factors.
At the hadron scale, the pressure expresses forces of attraction/repulsion and shear stress in the bound-state formed by the quasiparticle constituents.  Here there is a direct analogy with realisable two- and three-body systems in quantum mechanics.
At higher resolving scales, $\zeta$, reached via evolution, one might view a hadron interior as a dense medium of partons.  In this case, the pressure expresses pairwise forces between test elements within the medium. Each such element contains a number of partons determined by its volume and the associated species light-front wave function (LFWF) magnitude-squared at its location.
In all these things, the pressure distribution form factor and its interpretations are qualitatively equivalent to the distributions of charge, magnetisation, mass, spin, etc.





Today it is common to give some attention to a form factor that may be associated with the in-proton expectation value of the QCD trace anomaly:
\begin{equation}
\theta(Q^2) =
A(Q^2) + \frac{Q^2}{4 m_p^2} [A(Q^2)-2 J(Q^2) + 3 D(Q^2)]\,.
\label{theta}
\end{equation}
(The only difference from Eq.\,\eqref{MQ2} is $D \to 3 D$.)
Again, as a property of the proton itself and being a Poincar\'e-invariant form factor, the function is an objective result.
On the other hand, any interpretation of the result in terms of one or another set of underlying/embedded/sublimated degrees of freedom is practitioner and scale dependent.
For instance, any statement that $\theta(Q^2)$ expresses something about gluon contributions to the nucleon mass is subjective.  The definition of glue is scale dependent and no probabilistic interpretation is possible in anything except the infinite momentum frame.  Further, long wavelength observables, such as the total nucleon mass, have no objective species decomposition.

\begin{figure}[t]
\includegraphics[clip, width=0.485\linewidth]{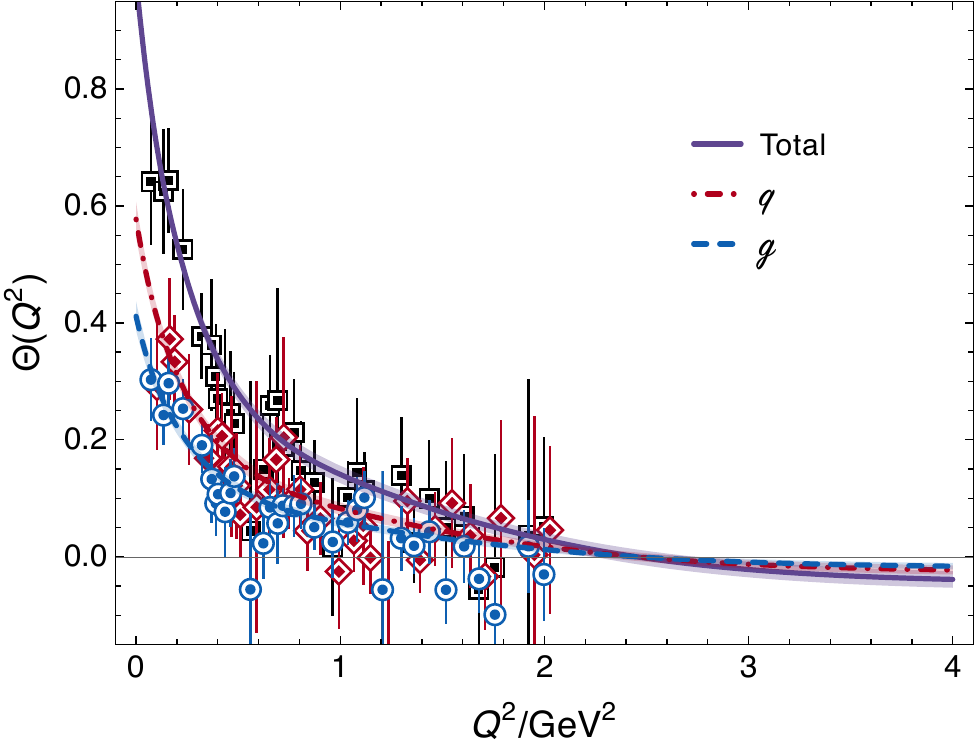} \hspace*{\fill}
\includegraphics[clip, width=0.485\linewidth]{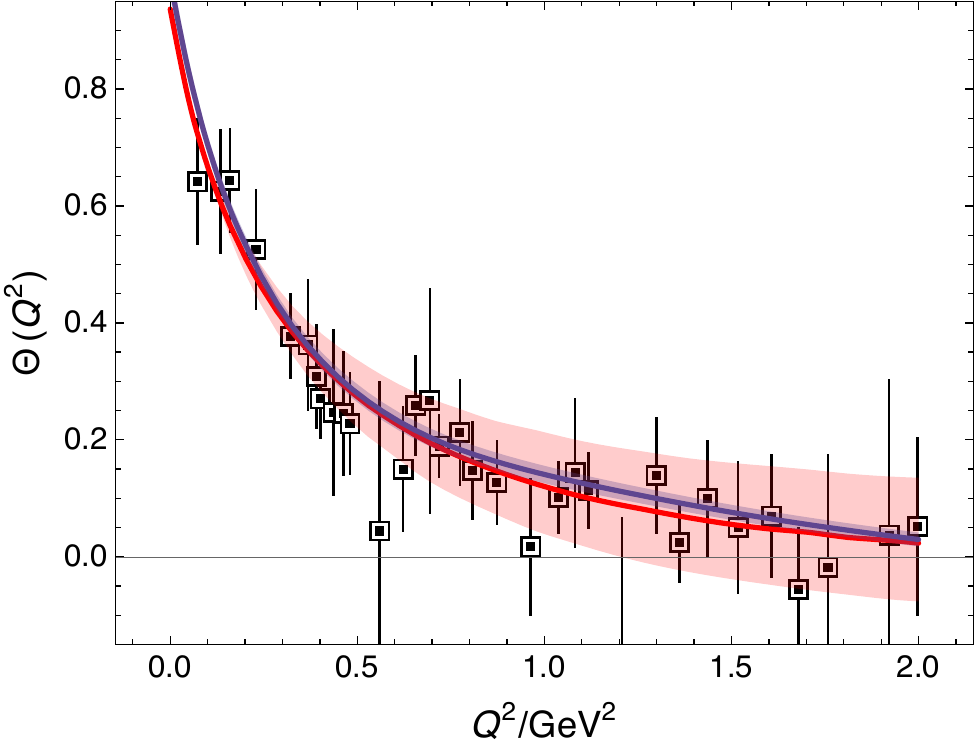}
\caption{\label{thetaQ2}
\emph{Left panel}.
    Curves -- CSM predictions: bracketing bands mark the extent of $1\sigma$ uncertainty in the numerical procedure for reaching large $Q^2$.
    In each case, the overall (species-summed) result is independent of resolving scale, $\zeta$.
    The species decompositions evolve with $\zeta$.
    The lQCD points in each panel are obtained using the results in Ref.\,\cite{Hackett:2023rif}:
    black squares -- total form factor;
    red diamonds -- quark component;
    blue circles -- glue.
\emph{Right panel}.
    $\theta(Q^2)$.
    Purple curve and band -- CSM prediction.
    Red curve and band -- data-based inference in Ref.\,\cite{Cao:2024zlf}.
}
\end{figure}

Figure~\ref{thetaQ2}-left depicts the CSM prediction for $\theta(Q^2)$ along with the species separation delivered by AO evolution \cite{Yao:2024ixu}.
Evidently, within the large lattice uncertainties, there is agreement between lQCD and the CSM results in every case.

Using the CSM results displayed in Fig.\,\ref{thetaQ2}-left, one finds
\begin{equation}
    \langle r^2 \rangle_{\theta}=
    (0.960(36)\,{\rm fm})^2 = (1.08(4) r_{\rm ch})^2\,.
    \label{rtheta2}
\end{equation}
This prediction compares favourably with the data-informed extraction in Ref.\,\cite{Cao:2024zlf}: $(0.97^{+0.02}_{-0.03}\,{\rm fm})^2$.

The objective interpretation of Eq.\,\eqref{rtheta2} is that one particular combination of proton gravitational form factors has a larger slope than another chosen combination of electromagnetic form factors.
There is no sense in which this means that gluons are distributed over a larger spacetime volume than quarks because the total form factor, being scale invariant, does not express information about a subjective species decomposition.
This is highlighted by the fact that the result in Eq.\,\eqref{rtheta2} is the $\theta(Q^2)$ radius associated with the valence quark quasiparticles that feature in the solution of the Poincar\'e-covariant Faddeev equation, into which all glue contributions are sublimated.
A species decomposition is provided in Fig.\,\ref{thetaQ2}-left; and as with ${\cal M}(Q^2)$, at a scale of $\zeta = 2\,$GeV, the quark contribution to $\langle r^2 \rangle_{\theta}$ is $\approx 40$\% larger than the gluon contribution: $\langle r^2 \rangle_{\theta}^{\mathscr q} \approx 1.4 \langle r^2 \rangle_{\theta}^{\mathscr g}$.

In Fig.\,\ref{thetaQ2}-right, we display a comparison between our prediction for $\theta(Q^2)$ and that presented in Ref.\,\cite{Cao:2024zlf}.  Evidently, the agreement is so good that the two distinct curves match almost within line-width on the displayed domain.

A disturbing feature, highlighted by Fig.\,\ref{thetaQ2}-right, is that $\theta(Q^2)$ is not positive definite.
Figure~Fig.\,\ref{thetaQ2}-left shows that the species decompositions have the same feature.
This is an unusual outcome to be associated with a global property of a nucleon mass distribution and equally peculiar for the species decompositions.
Whilst distributions of charge involve positive and negative charge carriers, hence can possess domains of negative support; all physical mass is positive definite, so there is seemingly no good reason why a form factor definitive of mass should become negative.
The mass-energy form factor in Eq.\,\eqref{MQ2} is positive definite.
The complication arises because of the change $1 D(Q^2) \to 3 D(Q^2)$ in going from Eq.\,\eqref{MQ2} to Eq.\,\eqref{theta}.

The analysis in Ref.\,\cite{Yao:2024ixu} included a species decomposition of the pressure form factor; and focusing on light quarks alone, it delivered $D^{u+d}(0;\zeta_2) = -1.73(5)$.
For comparison, a data-based inference yields $D^{u+d}(0) = -1.63(29)$ \cite{Burkert:2018bqq}.
Moreover, within uncertainties, the CSM prediction is in $Q^2$ pointwise agreement with the extraction therein -- see Ref.\,\cite[Fig.\,2]{Yao:2024ixu}.

It should be remarked that nothing similar can be said about the gluon contribution to $M(Q^2)$, $\theta(Q^2)$.  Whilst it was long hoped that near-threshold $J/\psi$ photoproduction could provide such information, this is now known to be forlorn because the reaction models underlying this connection are unrealistic.  Many modern analyses of $J/\psi$ photoproduction can provide an excellent description of available data without reference to any property intrinsic to the proton -- see, \emph{e.g}., Refs.\,\cite{Du:2020bqj, JointPhysicsAnalysisCenter:2023qgg, Tang:2024pky, Sakinah:2024cza}.

\section{Nucleon Resonances}
\label{SecResonance}
The quark model spectroscopic labelling convention for hadrons is still popular, \emph{viz}.\ cataloguing hadrons as $n\,^{2s+1}\!\ell_J$ systems, where $n$, $s$, $\ell$ are radial, spin, and orbital angular momentum quantum numbers, with $\ell+s=J$.  However, thinking deeper about hadron structure, Poincar\'e covariance entails that every hadron contains orbital angular momentum.  For instance, even the pion contains two {\sf S}-wave and two {\sf P}-wave components \cite{Bhagwat:2006xi, Krassnigg:2009zh}.
Moreover, no system is simply a radial excitation of another: there are simply too many degrees of freedom for this to be the case.

Adding to these things, whilst the total angular momentum, $J$, is Poincar\'e invariant -- so, observable -- any separation of $J$ into $L+S$ is frame dependent, \emph{i.e}., subjective.
Consequently, \emph{e.g}., negative parity states are not merely orbital angular momentum excitations of positive parity ground states.
Furthermore, in quantum field theory, parity and orbital angular momentum are unconnected.  This is obvious because parity is a Poincar\'e invariant quantum number, whereas $L$ is not; and no subjective quantity can properly define an observable.

It should now be clear that the QCD structure of hadrons, both mesons and baryons, is far richer than can be produced by quark models.  Hence, given that baryons are the most fundamental three-body systems in Nature, then to understand Nature we must understand how QCD, a Poincar\'e-invariant quantum field theory, builds each of the baryons in the complete spectrum.

In connection with this goal, we have illustrated above that the prediction of nucleon properties using matrix elements built with elementary Schwinger functions is now possible.  Notwithstanding that, it remains a challenging task.  So for many applications, a quark + diquark picture of baryons is used because it greatly simplifies the problem \cite{Barabanov:2020jvn}.  In modern analyses, the diquark correlations are nonpointlike; fully dynamical, \textit{viz}.\ subject to continual breakup and reformation and interacting with all quark-sensitive probes; and characterised by mass scales that are heavier than their natural mesonic analogues.  It is worth remarking that owing to the character of SU$(3)$-color, baryons are the most likely systems within which diquarks might play a dominant role.
In four- and five-body systems, \textit{i.e}., the putative tetra- and penta-quarks, (molecule-like) composites of color singlet states can exist.
They are typically more likely because diquark subcomponents would be heavier; hence, diquark subclusters would only be favored in exceptional circumstances -- see, \textit{e.g}., Ref.\,\cite{Hoffer:2024alv}.

\begin{figure}[t]
\includegraphics[clip, width=0.485\linewidth]{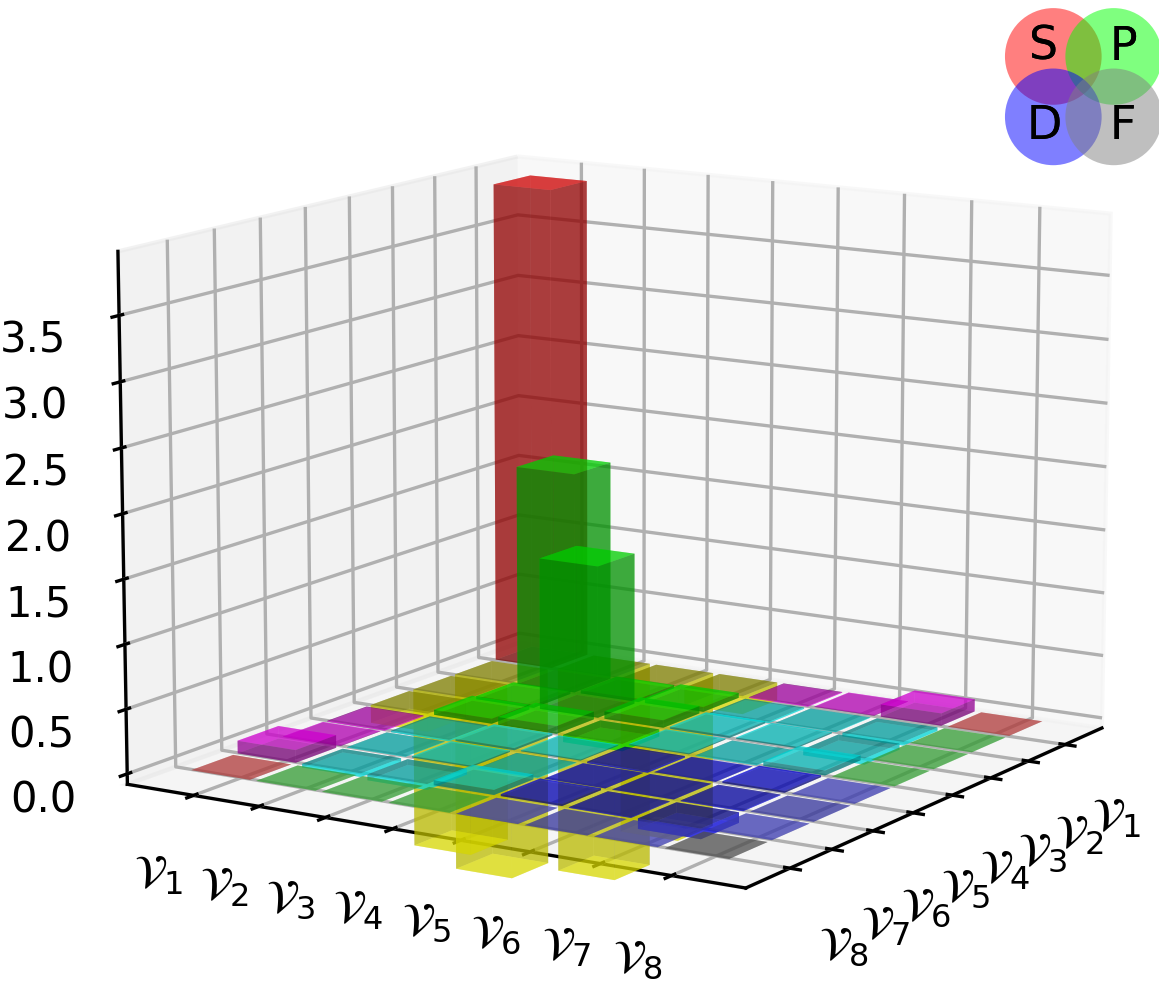} \hspace*{\fill}
\includegraphics[clip, width=0.485\linewidth]{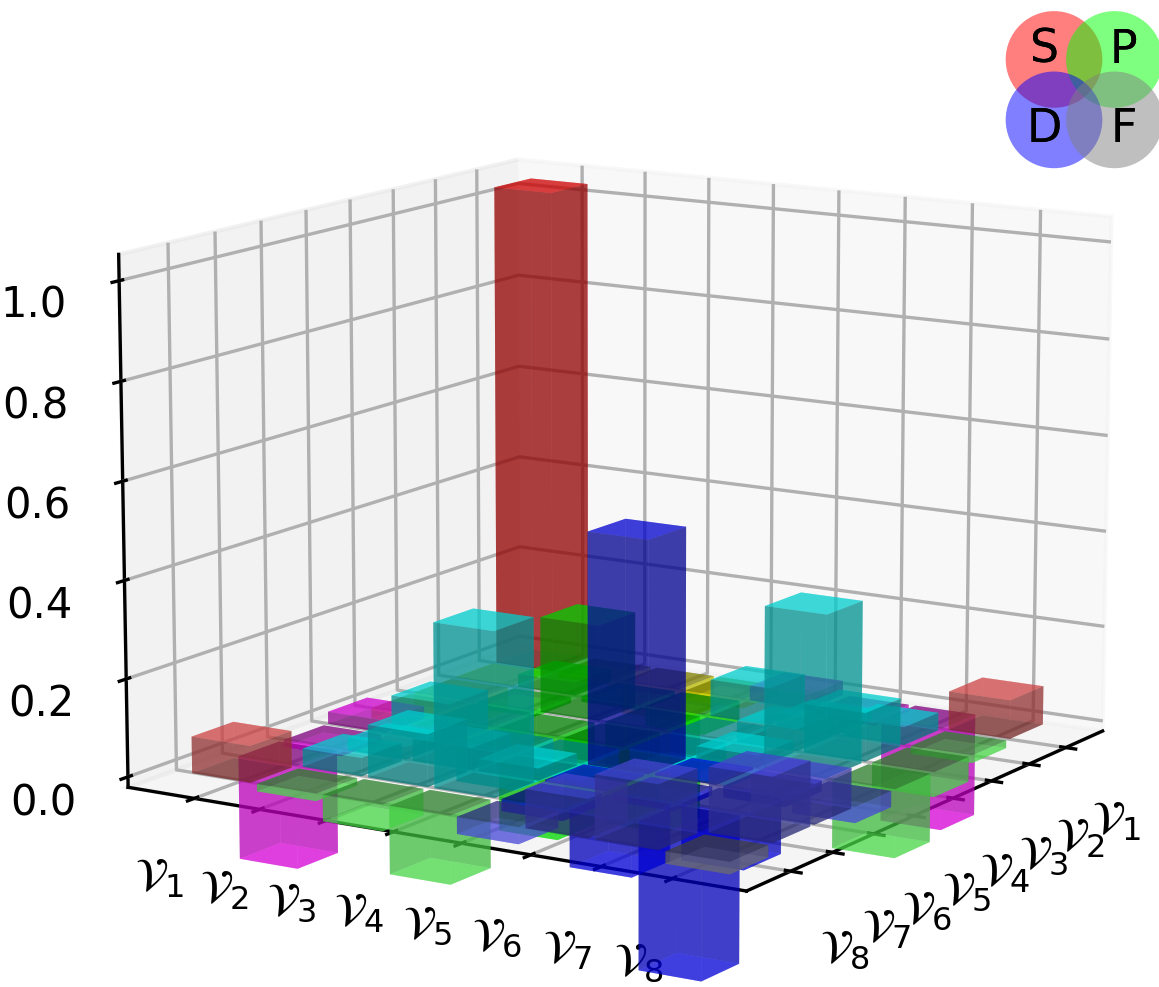}
\caption{\label{FOrbital}
Rest frame quark + axialvector diquark orbital angular momentum content of the $\Delta(1232)$ (left panel) and $\Delta(1600)$ (right panel), as measured by the contribution of the various components to the associated canonical normalization constant.
(See Ref.\,\cite{Liu:2022ndb} for details.)
}
\end{figure}

In developing the quark + diquark approximation to the baryon problem, one finds that both scalar and axialvector diquark correlations are present in the nucleon as a consequence of EHM -- see, \emph{e.g}., Ref.\,\cite[Fig.\,2]{Liu:2022ndb}.  Moreover, basically because one is dealing with a bound state in relativistic quantum field theory, as highlighted above, all baryon wave functions contain significant orbital angular momentum.  This is illustrated in Fig.\,\ref{FOrbital}, which displays rest frame partial wave decompositions of the Poincar\'e-covariant wave functions for the $\Delta(1232)$ and $\Delta(1600)$.  (Since these are isospin $3/2$ baryons, they only contain axialvector-isovector diquarks.)

Studying the Poincar\'e covariant wave functions of the $\Delta(1232)$ and $\Delta(1600)$, one is led to conclude that these systems are related as ground state and (principally) first radial excitation \cite{Liu:2022ndb}.
Namely, the same $(I)J^P=(\tfrac{3}{2})\tfrac{3}{2}^+$ bound-state equation delivers separated solutions, where the wave function of the lowest mass solution, $\Delta(1232)$, can be characterised as possessing no zeros whilst that of the more massive solution, $\Delta(1600)$, displays a single zero.

Focusing first on Fig.\,\ref{FOrbital}-left, one sees that the canonical normalisation of the $\Delta(1232)$ is principally determined by $\mathsf S$-wave components, but there are large, constructive $\mathsf P$ wave contributions and also strong $\mathsf S\otimes \mathsf P$-wave destructive interference terms (off-diagonal, negative-value yellow bars).  This picture of the $\Delta(1232)$ structure has been confirmed by comparisons with data on the $\gamma+p \to \Delta(1232)$ transition form factors \cite{Eichmann:2011aa, Segovia:2014aza, Lu:2019bjs}.

Shifting attention to Fig.\,\ref{FOrbital}-right, $\mathsf S$-wave contributions are dominant in the $\Delta(1600)$, too, but there are prominent $\mathsf D$-wave components, material $\mathsf P \otimes \mathsf D$-wave interference contributions (off-diagonal cyan blocks), and numerous $\mathsf F$-wave induced interference terms (off-diagonal blocks near ${\cal V}_8 {\cal V}_8$).
Enhanced higher partial waves are also seen in related three-body Faddeev equation studies of the $\Delta(1600)$ \cite{Eichmann:2016hgl, Qin:2018dqp}.
This quark+diquark structural picture of the $\Delta(1600)$, which represents it as the first radial excitation of the $\Delta(1232)$, albeit with significant orbital angular momentum excitations as well, was used to calculate $\gamma^\ast+p \to \Delta(1600)$ transition form factors \cite{Lu:2019bjs}.  Those predictions were confirmed in analyses of $\pi^+ \pi^- p$ electroproduction data collected at JLab \cite{Mokeev:2023zhq}.

It is likely that any approach to calculating $\Delta(1600)$ properties which fails to express its character as a combined radial and orbital angular momentum excitation of the $\Delta(1232)$, with a truly complex wave function, including the orbital angular momentum structure in Fig.\,\ref{FOrbital}-right, will deliver an erroneous structural picture of this state.  Fortunately, now that electroexcitation data exist \cite{Mokeev:2023zhq}, proponents of alternative interpretations of the $\Delta(1600)$ can subject their pictures to stringent validation tests.  

\section{Parton Distribution Functions}
\begin{figure}[t]
\includegraphics[clip, width=0.485\linewidth]{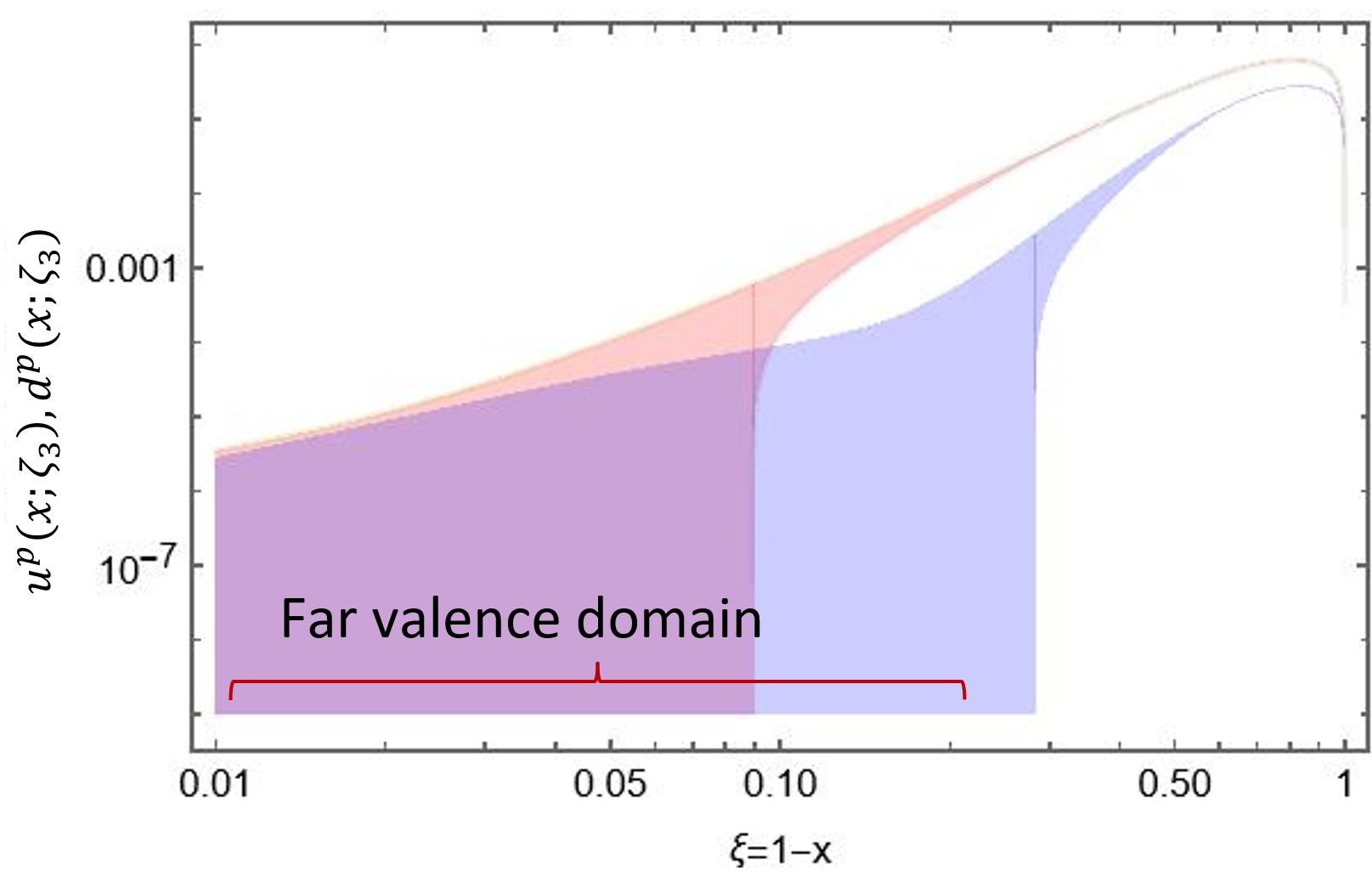} \hspace*{\fill}
\includegraphics[clip, width=0.485\linewidth]{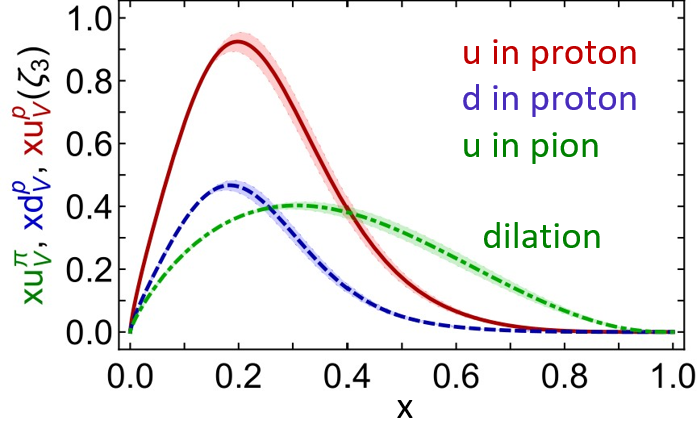}
\caption{\label{FDFs}
\emph{Left panel}.  Proton valence quark DFs inferred from data as described in Ref.\,\cite[NNPDF]{NNPDF:2021njg}.
\emph{Right panel}.  CSM predictions for proton and pion dressed valence quark DFs \cite{Lu:2022cjx}.
(In both panels, $\zeta_3 = 3\,$GeV.)
}
\end{figure}

Despite many years of effort, much must still be learnt before proton and pion structure may be considered understood in terms of parton distribution functions (DFs).
One illustration of this problem is provided in Fig.\,\ref{FDFs}-left, which displays $\zeta=\zeta_3:= 3.1\,$GeV inferences of proton valence quark DFs by the NNPDF Collaboration \cite{NNPDF:2021njg}.  The abscissa is $\xi=1-x$, where $x$ is the light-front fraction of the proton momentum carried by the identified quark.  Evidently, insofar as one may determine from phenomenological fitting of data, nothing is known about these DFs on the valence domain, $\xi\lesssim 0.3 \leftrightarrow x \gtrsim 0.7$.  The fits even permit the DFs to become negative, which, given the LFWF overlap representation of DFs \cite{Diehl:2003ny}, is physically impossible for an unpolarised DF.

The basic question is simple: Given, \emph{e.g}., the marked disparities between masses and mass budgets -- see Fig.\,\ref{MassBudgets}, what are the differences, if any, between the distributions of partons within the proton and the pion?
This question has special resonance today as science seeks to explain EHM.

CSM predictions \cite{Lu:2022cjx} for proton and pion dressed valence-quark DFs are drawn in Fig.\,\ref{FDFs}-right, with the resolving scale matching that in the left panel.  These DFs are positive definite on $x\in (0,1)$ and satisfy the endpoint constraints anticipated from QCD.  Namely, considering valence DFs associated with processes that do not involve beam or target polarisation, then \cite{Brodsky:1994kg, Yuan:2003fs, Cui:2021mom, Cui:2022bxn}:
\begin{equation}
{\mathscr d}^p(x;\zeta_{\cal H}), {\mathscr u}^p(x;\zeta_{\cal H}) \stackrel{x\simeq 1}{\propto} (1-x)^{3+\gamma_p(\zeta)}\,,\quad
\bar {\mathscr d}^\pi(x;\zeta_{\cal H}), {\mathscr u}^\pi(x;\zeta_{\cal H})  \stackrel{x\simeq 1}{\propto} (1-x)^{2+\gamma_\pi(\zeta)}\,;\;
\label{LargeX}
\end{equation}
where the anomalous dimensions $\gamma_{p,\pi}$ increase from zero with increasing $\zeta>\zeta_{\cal H}$ \cite{Dokshitzer:1977sg, Gribov:1971zn, Lipatov:1974qm, Altarelli:1977zs}.  The large $x$ exponent on the associated gluon DFs is roughly one unit larger; and that for the sea quark DFs is approximately two units larger -- see, \emph{e.g}., Ref.\,\cite[Table~1]{Lu:2022cjx}.

Regarding Fig.\,\ref{FDFs}-right, one sees that the in-pion valence quark DF is far more dilated than its proton analogues.  This owes partly to the different number of valence degrees of freedom in the proton and pion.  However, there is also an EHM-driven aspect to this dilation, which is expressed in the Goldberger-Treiman relation that connects the pion bound-state amplitude with the dressed light-quark mass function \cite{Maris:1997hd, Qin:2014vya}.   This physics drives pion DFs to be the most dilated in Nature.

Less obvious in Fig.\,\ref{FDFs}-right are the following two facts:
(\emph{i}) ${\mathscr u}^p(x) \neq 2 {\mathscr d}^p(x)$; and
(\emph{ii}) ${\mathscr d}^p(x) \propto {\mathscr u}^p(x)$ on $x\simeq 1$.
The first of these observations indicates that the proton wave function involves correlations which distinguish between $u$ and $d$ quarks.  This is readily seen via the quark + diquark picture, in which $[u,d]$ scalar-isoscalar diquarks sequester the $d$ quark into a soft correlation and the $d$ quark can only participate in hard interactions owing to the presence of $\{uu\}$ axialvector-isovector correlations.
The second observation highlights that both $u$ and $d$ quarks are valence degrees of freedom in the proton.  Here, the constant of proportionality is determined by the relative strength of scalar and axialvector diquarks in the proton wave function.  Modern quark + diquark descriptions of proton structure predict that axialvector diquarks are responsible for $\approx 40$\% of the proton's wave function normalisation \cite[Fig.\,2]{Liu:2022ndb}.  This level of contribution delivers a pointwise good description of available data on the ratio of neutron and proton $F_2$ structure functions \cite[BoNuS]{CLAS:2014jvt}, \cite[MARATHON]{Abrams:2021xum} -- see, \emph{e.g}., Ref.\,\cite[Fig.\,4B]{Lu:2022cjx}, with $\lim_{x\to 1} F_2^n(x)/F_2^p(x) = 0.453(46)$.  An objective SPM analysis of MARATHON data yields \cite{Cui:2021gzg}: $\lim_{x\to 1} F_2^n(x)/F_2^p(x) = 0.437(85)$.   If one were to eliminate axialvector diquark correlations from the nucleon wave function, then the result for this ratio would be $0.25$ \cite{Close:1988br}.
(Scalar diquark breakup and recombination effects do not change materially affect result \cite{Xu:2015kta}.)

Many other topical issues relating to pion and proton DFs are receiving insightful CSM treatments.  For instance, SeaQuest Collaboration data \cite{SeaQuest:2021zxb} on the asymmetry of antimatter in the proton are discussed in Refs.\,\cite{Lu:2022cjx, Chang:2022jri} and the proton spin puzzle in Refs.\,\cite{Cheng:2023kmt, Yu:2024ovn}.

\section{Conclusion}
The past decade has seen material progress in strong interaction theory.  Today, there is an expanding array of parameter-free predictions for the proton and, importantly, for many of the other hadrons whose properties express the full meaning of QCD.  For instance, insights are being drawn into the structure of Nature's most fundamental Nambu-Goldstone bosons -- pions and kaons -- and that of nucleon excited states and resonances, in this latter case capitalising on accumulating resonance electroproduction data.

A compelling motivation for all these efforts is the need to understand how QCD's Lagrangian simplicity can explain the emergence of the diverse and complex body of detectable hadronic states.  One may safely expect that the precise data needed to test any conjecture will become available during operations of modern high-luminosity, high-energy facilities.  A valid explanation will move science into a new realm of understanding by, perhaps, proving QCD to be the first well-defined four-dimensional quantum field theory ever contemplated.  If such is the case with QCD, then doors may open that lead far beyond the Standard Model.

\section*{Acknowledgments}
Work supported by:
National Natural Science Foundation of China (grant no.\ 12135007);
and
Spanish Ministry of Science, Innovation and Universities (MICIU grant no.\ PID2022-140440NB-C22).




\end{document}